\documentclass[twocolumn,aps,prl,10pt,amsmath,amssymb,nofootinbib,showpacs,superscriptaddress,floatfix]{revtex4-1}

\newcommand{\rs}{\rm\scriptscriptstyle}

\DeclareFontFamily{U}{rcjhbltx}{}
\DeclareFontShape{U}{rcjhbltx}{m}{n}{<->rcjhbltx}{}
\DeclareSymbolFont{hebrewletters}{U}{rcjhbltx}{m}{n}

\usepackage{graphicx}
\usepackage{color}
\usepackage[usenames,dvipsnames]{xcolor}
\usepackage[colorlinks=true,linkcolor=Red,citecolor=Green,linktoc=page]{hyperref}
\usepackage{multirow}
\usepackage{float}
\usepackage{flushend}
\usepackage{balance}
\usepackage[varg]{txfonts}
\usepackage{ulem}
\usepackage{fancyhdr}

\DeclareMathSymbol{\lamed}{\mathord}{hebrewletters}{108}

\begin{document}
\title{Confinement and asymptotic freedom with Cooper pairs}

\author{M.\,C.\,Diamantini}

\affiliation{NiPS Laboratory, INFN and Dipartimento di Fisica e Geologia, University of Perugia, via A. Pascoli, I-06100 Perugia, Italy}

\author{C.\,A.\,Trugenberger}

\affiliation{SwissScientific Technologies SA, rue du Rhone 59, CH-1204 Geneva, Switzerland}

\author{V. M. Vinokur}
\affiliation{Materials Science Division, Argonne National Laboratory, 9700
S. Cass Ave., Lemont, IL 60439, USA}

\begin{abstract}
	
One of the most profound aspects of the standard model of particle physics, the 
mechanism of confinement binding quarks into hadrons, is not sufficiently understood. The only known semiclassical mechanism of confinement, mediated by chromo-electric
strings in a condensate of magnetic monopoles still lacks experimental evidence.
Here we show that the infinite
resistance superinsulating state, which emerges on the insulating
side of the superconductor-insulator transition in superconducting films
offers a realization of confinement that allows for a direct experimental access. 
We find that superinsulators realize a single-color version of quantum chromodynamics and establish the mapping of quarks onto Cooper pairs. We reveal that the mechanism of superinsulation is the linear binding of Cooper pairs into neutral ``mesons" by electric strings. 
Our findings offer a powerful laboratory
for exploring and testing the fundamental implications of confinement, asymptotic
freedom, and related quantum chromodynamics phenomena
via the desktop experiments on superconductors.
\end{abstract}
\maketitle


\section*{Introduction}
The standard model of particle physics is extraordinarily successful at explaining many facets of the physical realm. 
Yet, one of its 
profound aspects, the mechanism of confinement binding quarks
into hadrons, is  not sufficiently understood. 
The only known semiclassical mechanism of confinement is mediated by chromo-electric
strings in a condensate of magnetic monopoles\,\cite{mandelstam, thooft, polyakov_original} but its relevance for quantum chromodynamics still lacks experimental evidence. 
This suggests a quest for systems that could allow for direct experimental tests of the string confinement mechanism. To identify such a system we follow a brilliant insight of `t Hooft\,\cite{thooft1978},  who appealed to a solid state physics analogy in a Gedankenexperiment to explain quark confinement. He demonstrated that it is realized in a phase which is a dual twin to  superconductivity, in a sense that it has zero particle mobility, and called hence this phase a ``superinsulator." 
The infinite-resistance superinsulating state was indeed first predicted to emerge in Josephson junction arrays (JJA)\,\cite{Diamantini1996} and then in disordered superconducting films\,\cite{Doniach1998,vinokur2008superinsulator} at the insulating side of the superconductor-insulator transition (SIT)\,\cite{Efetov1980,Haviland1989,Paalanen1990,Fisher1990,fazio}.
Experimentally, superinsulators were observed in titanium nitride (TiN) film\,\cite{vinokur2008superinsulator,Baturina2007} and, albeit under different name, InO films\,\cite{Shahar2005} and have become ever since a subject of an intense study, see\,\cite{vinokurAnnals,Shahar,Mironov2018} and references therein.

Originally, the idea of superinsulation\,\cite{Diamantini1996,vinokur2008superinsulator} grew from the supposed 2D logarithmic Coulomb interactions between Cooper pairs in the critical vicinity of the SIT realized in lateral Josephson junction arrays\,\cite{fazio,Diamantini1996}. 
Here we show that, starting with the uncertainty principle for Cooper pairs\,\cite{vinokur2008superinsulator} and building solely on the most general locality and gauge invariance principles, one constructs the effective action for superinsulators, which is exactly
Polyakov's compact quantum electrodynamic (QED) action\,\cite{polyakov_original, polyakov}.  Accordingly, superinsulation emerges as an explicit realization of the Mandelstam --`t\,Hooft S-duality\,\cite{mandelstam,thooft} in materials that harbor Cooper pairs and constitutes a single-color version of the quantum chromodynamic (QCD) vacuum, in which Cooper pairs play the role of quarks. We thus find that the Cooper pair binding mechanism in a superinsulator, leading to the infinite resistance at finite temperatures, is the linear, rather than logarithmic, confinement of charges into neutral ``mesons" due to Polyakov's electric strings\,\cite{polyakov_original, polyakov}, arising in the vortex condensate.  The Abelian character of the compact QED, albeit a strong coupling gauge theory, allows for an analytical derivation of the linear confinement by electric strings, at variance to the QCD whose complexity requires heavy numerical computations. 

\begin{figure}[t!]
	\includegraphics[width=9cm]{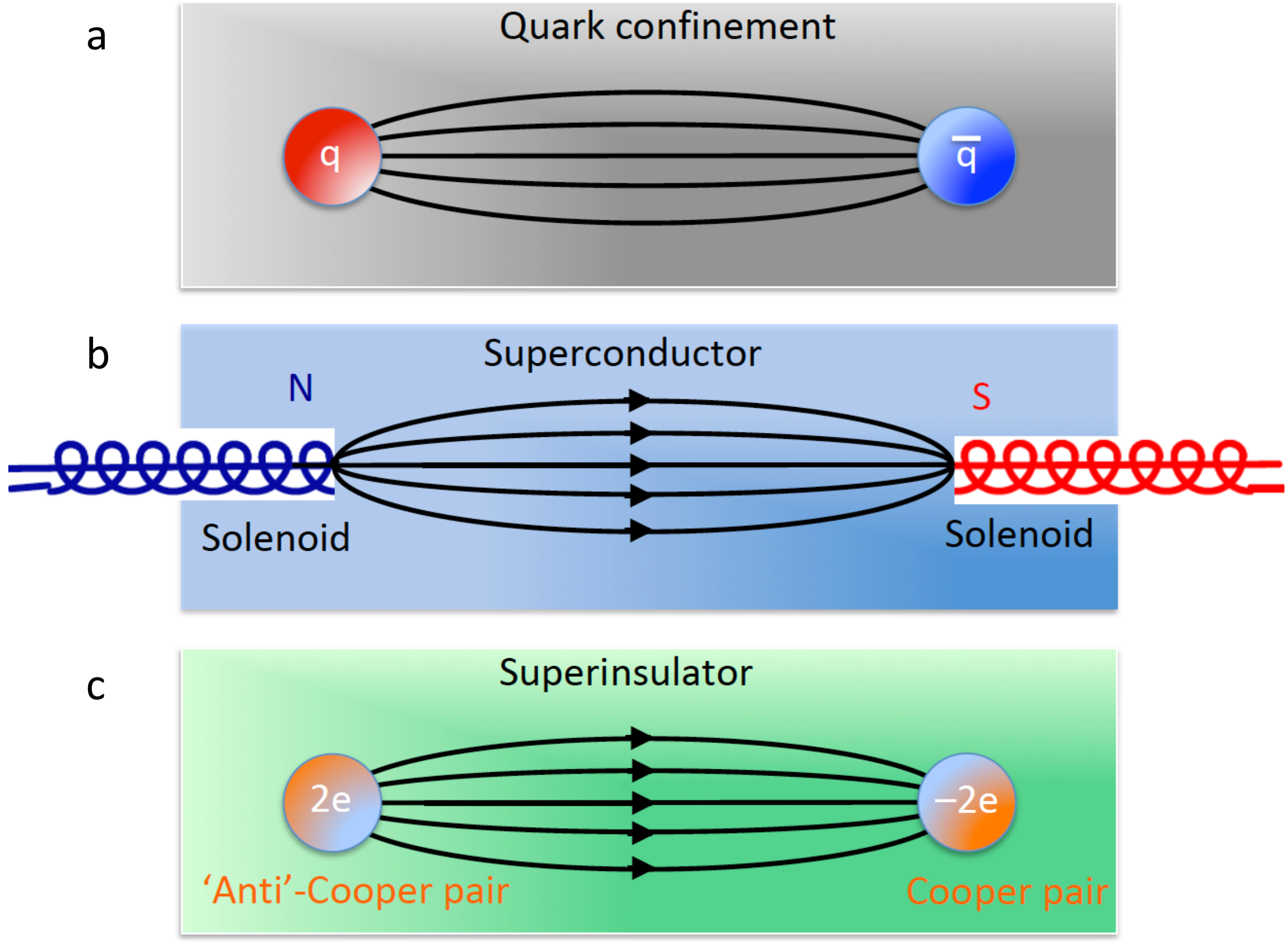}
	\caption{\textbf{Dual Mandelstam--`t\,Hooft--Polyakov confinement.} \textbf{a}:\,quark confinement by chromo-electric strings. \textbf{b}:\,magnetic tube (Abrikosov vortex) that forms in a superconductor between two magnetic monopoles. \textbf{c}:\,electric string that forms in a superinsulator between the Cooper pair and anti-Cooper pair. The lines are the force lines for magnetic and electric fields respectively. In all cases the energy of the string (the binding energy) is proportional to the distance between either the monopoles or the charges.}
	\label{fig:Fig.1} 
\end{figure}

Since linear confinement by strings is not restricted to 2D, we establish that superinsulation is a distinct genuine state of matter that appears in both 2D or 3D realizations  
and calculate the deconfinement temperature that marks the phase transition of superinsulators into conventional insulators and which, in 2D, coincides with the Berezinskii-Kosterlitz-Thouless (BKT) transition temperature. 
Finally we also unearth a Cooper pair analogue of the asymptotic freedom effect\,\cite{gross}, which suggests that systems smaller than the string scale appear in a quantum metallic state. 
Our findings offer thus an easy access tool for testing fundamental implications of confinement, asymptotic freedom, and related QCD phenomena via desktop experiments on superconductors. 

\section*{Results}
\subsection{Action in two-dimensional systems}
We start  by showing how dual superconducting and superinsulating states can be understood from the uncertainty principle, $\Delta N\Delta\varphi\geqslant 1$ between the number of charges, $N=2|\Psi|^2$, and the phase $\varphi$ of the Cooper pairs quantum field $\Psi=N\exp(i\varphi)$, bound by the commutation relation $[N,\varphi]=i$\,\cite{vinokurAnnals,dtlv}. At zero temperature, superconductors correspond to fixed $\varphi$, hence indefinite $N$. Inversely, fixed $N$ and indefinite $\varphi$ characterizes the superinsulating state. As a Cooper pair is a charge quantum, while a vortex carries the $2\pi$ phase quantum, the SIT is driven by the competition between charge (Cooper pairs) and vortex degrees of freedom, in accordance with early ideas\,\cite{Fisher1990}. 

We turn now to the construction of the action of the Cooper pair-vortex system near the SIT, where both degrees of freedom are to be included on an equal footing. The key contribution is the infinite-range (i.e. non-decaying with distance) Aharonov-Bohm-Casher (ABC) Cooper pair-vortex topological interaction, embodying the quantum phase acquired either by a charge encircling a vortex or by a vortex encircling a charge. 
To ensure a local formulation of the action, we must introduce two emergent gauge fields, $a_\mu$ and $b_\mu$ mediating these ABC interactions. Then the topological part of the action assumes the form
\begin{equation} 
S^{\rs CS} =  \int d^3x \left[\ i  {n \over 2\pi} a_{\mu} \epsilon_{\mu \alpha \nu} \partial_{\alpha} b_{\nu} + i\sqrt{n} a_{\mu} Q_{\mu} +i\sqrt{n} b_{\mu} M_{\mu}\right]\,,
\label{CS}
\end{equation}
where $\epsilon_{\mu \alpha \nu}$ is the completely antisymmetric tensor, and
\begin{eqnarray}
Q_{\mu} = \sum_{\mathrm i} \int_{\rm x_{q}^{(i)}} d\tau {dx_{{\mathrm q}\mu}^{({ i})}(\tau) \over d\tau} \ \delta^3 \left( x- x_{\mathrm q}^{({ i})}(\tau) \right) \ ,
\nonumber \\
M_{\mu} = \sum_{\mathrm i} \int_{\rm x_{m}^{(i)}} d\tau {dx_{{\mathrm m}\mu}^{({ i})}(\tau) \over d\tau} \ \delta^3 \left( {x}- {x_{\mathrm m}^{(i)}(\tau)} \right) \ ,
\label{topex2D}
\end{eqnarray}
are the world-lines of elementary charges and vortices labeled by the index ${\mathrm i}$, parametrized by the coordinates $x_{\mathrm q}^{(i)}$ and $x_{\mathrm m}^{(i)}$, respectively, 
$n$ is the dimensionless charge, and Greek subscripts run over the Euclidean three dimensional space encompassing the 2D space coordinates and the Wick rotated time coordinate. Equation\,(\ref{CS}) defines the mixed Chern-Simons (CS) action\,\cite{jackiw} and represents the local formulation of the topological interactions between charges and vortices, where the ABC phases are encoded in the Gauss linking number of the $\{x_q^{(i)}\}$ and $\{x_m^{(i)}\}$  world-lines. The CS action is invariant under the gauge transformations $a_{\mu} \to a_{\mu} + \partial_{\mu} \lambda$ and  $b_{\mu} \to b_{\mu} + \partial_{\mu} \chi$, reflecting the conservation of the charge and vortex numbers and
is the dominant contribution to the action at long distances, since it contains only one field derivative.  In this representation $j_{\mu} =(\sqrt{n}/2\pi) \epsilon_{\mu \alpha \nu} \partial_{\alpha} b_{\nu}$ and $\phi_{\mu} =(\sqrt{n}/2\pi) \epsilon_{\mu \alpha \nu} \partial_{\alpha} a_{\nu}$ are the continuous charge and vortex number current fluctuations, while $Q_{\mu}$ and $M_{\mu}$ stand for integer point charges and vortices. We use natural units $c=1$, $\hbar = 1$ but restore physical units when necessary. Also, from now on we set the charge unit $n=2$ for Cooper pairs. 

The next-order terms in the effective action of the SIT contain two field derivatives. Gauge invariance requires that they be constructed in terms of the ``electric" and ``magnetic" fields corresponding to the two gauge fields. Introducing the dual field strengths $f_{\mu} = \epsilon_{\mu \alpha \nu} \partial_{\alpha} b_{\nu}$ and $g_{\mu} = \epsilon_{\mu \alpha \nu} \partial_{\alpha} a_{\nu}$ one identifies the magnetic fields as $f_0$ and $g_0$ and the electric fields as $f_i$ and $g_i$, where "0" denotes the Wick rotated time and Latin indices denote purely spatial components. We thus arrive at the full action
\begin{widetext}
\begin{eqnarray}
S_{\rs 2D}=\int d^3 x\  i  {1\over \pi} a_{\mu} \epsilon_{\mu \alpha \nu} \partial_{\alpha} b_{\nu} 
+ {1\over 2e^2_{\mathrm v} \mu_{\rs P}} f_0^2+ 
{\varepsilon_{\rs P} \over 2e^2_{\mathrm v}} f_i^2
+{1 \over 2e^2_{\mathrm q}\mu_{\rs P}} g_0^2 + {\varepsilon_{\rs P} \over 2e^2_{\mathrm q}} g_i^2
+i\sqrt{2} a_{\mu} Q_{\mu} +i\sqrt{2} b_{\mu} M_{\mu}\,.
\label{nonrel}
\end{eqnarray}
\end{widetext}
Here $\mu_{\rs P}$ is the magnetic permeability and $\varepsilon_{\rs P}$ is the electric permittivity\,\cite{dtlv}, which define the speed of light $v_{\mathrm c} = 1/\sqrt{\mu_{\rs P} \varepsilon_{\rs P}}$ in the material. The two coupling constants, $e_{\mathrm q}^2=e^2/d$ and $e_{\mathrm v}^2 = \pi^2 /(e^2\lambda_{\perp})$  are the characteristic energies of a charge and  a vortex in the film, respectively\,\cite{dtlv}. Here $d$ is the thickness of the film, $\lambda_{\perp} = \lambda_{\rs L}^2/d$ is the Pearl length, and $\lambda_{\rs L}$ is the London length of the bulk. 
The effective action in this order of the expansion with respect to derivatives  is perfectly dual under the mutual exchange of charge and vortex degrees of freedom and the corresponding coupling constants. The charge-vortex duality is expressed by the action symmetry with respect to the transformation $g\equiv e_{\mathrm v}/e_{\mathrm q}\leftrightarrow 1/g$. Thus $g$ is the tuning parameter driving the system across the SIT, and the SIT itself corresponds to $g=g_{\mathrm c}=1$. The possible duality breaking is a higher order effect. In field theory, this duality goes under the name of S-duality (strong-weak coupling duality). 
Note that the addition of kinetic terms generates the topological Chern-Simons mass $m_{\rs T}$ for both gauge fields. In the relativistic case, $\mu_{\rs P} = \varepsilon_{\rs P} = 1$, and the CS mass becomes $m_{\rs T}= e_{\mathrm q} e_{\mathrm v}/\pi$\,\cite{jackiw}. In the non-relativistic case the CS mass is modified to $m_{\rs T} = \mu_{\rs P}  e_{\mathrm q} e_{\mathrm v}/\pi$ and the dispersion relation becomes $E = \sqrt{m_{\rs T}^2 v_{\mathrm c}^4 + v_{\mathrm c}^2 p^2}$, see Methods, Lattice Chern-Simons operator. 
We stress here that we derived the action (\ref{nonrel}) describing the system of interacting Cooper pairs and vortices using solely symmetry and gauge invariance considerations. Importantly,  the action describing Josephson junction arrays\,\cite{fazio, Diamantini1996} is a special case of the same action with $\varepsilon_{\rs P}=1$, $\mu_{\rs P}\to\infty$, $e_{\mathrm q}\to 4E_{\rs C}$, $e_{\mathrm v}\to 2\pi^2 E_{\rs J}$, where
$E_{\rs C}$ and  $E_{\rs J}$ 
are the charging energy and the Josephson energy of a single junction, respectively, 
see Supplementary note 1, 
Gauge theory of JJA. 
This provides a crosscheck for our general result. 
\bigskip

%


\subsection{Superinsulator}
We are now equipped to discuss the nature of the superinsulating state. To that end,
we couple the charge current $j_{\mu}$ to the physical electromagnetic gauge field $A_{\mu}$ by adding to the action the minimal coupling term $2e A_{\mu} j_{\mu}$. Setting $Q_{\mu}=0$, since charges are dilute, integrating out the gauge fields $a_{\mu}$ and $b_{\mu}$, and summing over the condensed vortices $M_{\mu}$, we arrive at the effective action $S_{\rm eff}(A_{\mu})$ describing the electromagnetic response of an ensemble of charges in a superinsulator.  On a discretized lattice with spacing $\ell$, see Methods, Lattice Chern-Simons action, the effective action takes a form, See Supplementary note 2, Effective action for the superinsulator, in which one immediately recognizes a non-relativistic version of the Polyakov action for the compact QED model\,\cite{polyakov_original,polyakov}:
\begin{widetext} 
\begin{eqnarray}
S_{\rm eff}(A_{\mu})= S_{\rm compact }^{\rs 2D}
={\gamma^2 \over 2\pi^2}  \left\{\sum_{x} v_{\mathrm c} \left[ 1- {\rm cos} \left( 2e\ell^2 F_0 \right) \right]+ \sum_{x,i} {1\over v_{\mathrm c}} \left[ 1- {\rm cos} \left( 2e\ell^2 F_i \right) \right] \right\}.
\label{comp}
\end{eqnarray}
\end{widetext} 
Here the summation runs over the lattice grid $\{x\}$, $F_{\mu } = k_{\mu \nu} A_{\nu}$ is the dual electromagnetic field strength, $k_{\mu \nu}$ is the lattice 
Chern-Simons operator $\epsilon_{\mu \alpha \nu} \partial_{\alpha}$, see Methods, Lattice Chern-Simons operator, and 
$\gamma ^2= C\eta g/v_{\mathrm c}$ with $C$ being a numerical constant. 
The quantity $\eta= (1/\alpha) \lamed (\kappa,v_{\mathrm c}) 
\,$ characterizes the strength of quantum fluctuations, see Supplementary note 3, Quantum phase structure. Here $\kappa=\lambda_{\perp}/\xi$ is the Ginzburg-Landau parameter of the film, $\xi$ is the superconducting coherence length, taking on the role of the ultraviolet cutoff $\ell$, and, finally, $\alpha=e^2/(\hbar c)\approx 1/137$ is the fine structure constant. 

The physics of a superinsulator is governed by the spontaneous proliferation of instantons\,\cite{polyakov} $M=\partial_{\mu}  M_{\mu}$, corresponding to magnetic monopoles, so that the vortex number is not conserved in the vortex condensate. 
Then, in a mirror analogue to the formation of Abrikosov vortices in superconductors due to the Meissner effect mediated by the Cooper pair condensate, the magnetic monopole condensate constricts electric field lines connecting the charge-anticharge pair into electric strings\,\cite{polyakov_original,polyakov} confining Cooper pairs in superinsulators into ``mesons", see Fig.\,\ref{fig:Fig.1}. Indeed, as seen from the action\,(\ref{comp}), at large $\gamma$, the dynamical fields get squeezed into the vicinity of the paths minimizing the action, to form quantized fluxes $\ell^2 F_{\mu}$. 
The quantized electric flux tubes are the analogues of the strings mediating linear confinement of quarks into hadrons. Like Abrikosov vortices, for which the London penetration depth, the inverse of the Anderson-Higgs photon mass, sets the spatial scale of the decay of encircling supercurrents and magnetic field associated with the vortex, 
the characteristic lateral scale $w_{\rm string}$ for the decay of electric fields around the string is the inverse of the photon mass $m_{\mathrm{\gamma}}$\,\cite{caselle}, $w_{\mathrm{string}} = 1/(v_{\mathrm c} m_{\mathrm\gamma})$.
The typical ``meson" size instead, is given by the string tension $\sigma$. In the 2D relativistic ($v_c = 1$) model these are given by\,\cite{kogan}
\begin{eqnarray}
m_{\gamma} =   \frac{\gamma^2}{\sqrt{\pi} v_{\mathrm c} \ell} {\rm e}^{- \gamma^2/2\pi}  \ ,
\nonumber \\
\sigma_{\rs{2D}} =\frac {\pi^2m_{\gamma} v_{\mathrm c}^2} {4\ell \gamma^2} = \frac{\pi^{3/2}v_{\mathrm c} }{4 \ell^2} {\rm e}^{- \gamma^2/2\pi}  \ .
\label{tension}
\end{eqnarray}

\begin{figure}
	\includegraphics[width=9cm]{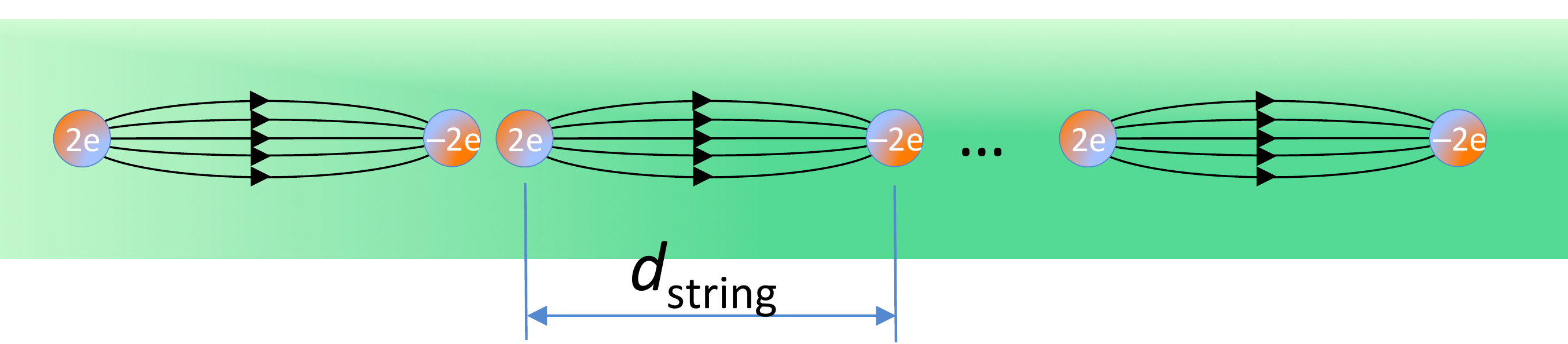}
	\caption{\label{fig:Fig4} \textbf{Splitting electric strings into neutral mesons.} The formation of a long string is energetically unfavorable, and small size charge-anticharge pairs emerge, splitting the string into a sequence of segments, each constituting a neutral meson.}
	\label{split}
\end{figure}

Unlike vortices, however, long strings are unstable: it is energetically favorable to break a string into a sequence of segments via the creation of charge-anticharge pairs, see Fig.\,(\ref{split}). This process corresponds to the creation of neutral ``mesons" with the typical size  $d_{\rm string} = \sqrt{v_{\mathrm c}/\sigma}$. From the dependence of $m_{\mathrm{\gamma}}$ and $\sigma$ on $\gamma^2$, one finds for the nonrelativistic case
\begin{equation}
d_{\rm string} \simeq \ell \ \exp\left( K {g\eta \over v^2_{\mathrm c}}\right)\,,
\label{nearsit}
\end{equation}
where $K$ is a numerical constant.
Near the SIT, where $g\approx 1/\eta$ and $v_{\mathrm c} =1/\sqrt{\mu_{\rs P} \varepsilon_{\rs P}} \ll c$ due to the divergence of the electric permittivity $\varepsilon_{\rs P}$\,\cite{vinokur2008superinsulator,vinokurAnnals}, $d_{\rm string}\gg\ell$, and the electric string is a well-defined object.
This establishes superinsulators as a single-color realization of QCD. Cooper pairs assume the role of quarks that are bound by electric strings into neutral mesons and this linear confinement is the origin of the infinite resistance of superinsulators. As quarks cannot be observed outside hadrons, Cooper pairs do not exists outside neutral bound states, and the absence of free charge carriers causes the infinite resistance.

\subsection{Action and superinsulator in three-dimensional systems}~~\\
The string confinement mechanism of superinsulation allows to generalize the concept of a superinsulator to higher dimensions, since linear confinement by electric strings is not specific to the 2D realm. Hence superinsulators can exist in 3D exactly as QCD exists in 3D. The 3D analogue of the topological action (\ref{nonrel}) involves the so called BF term\,\cite{bf}, combining the standard gauge field $a_{\mu}$ with the Kalb-Ramond antisymmetric gauge field of the second kind\,\cite{kalb} $b_{\mu \nu}$, 
\begin{widetext} 
\begin{eqnarray}
S_{\rs{3D}} = \int d^4x \ i  {1\over \pi} a_{\mu} \epsilon_{\mu \nu \alpha \beta} \partial_{\nu} b_{\alpha \beta} 
+{1\over 2e^2_{\mathrm v} \mu_{\rs P}} f_0^2+ {\varepsilon_P \over 2e^2_{\mathrm v}} f_i^2
+{1 \over 2e^2_{\mathrm q}\mu_{\rs P}} b_i^2+ {\varepsilon_{\rs P} \over 2e^2_{\mathrm q}} e_i^2
+i\sqrt{2} a_{\mu} Q_{\mu} +i{\sqrt{2}\over 2} b_{\mu \nu} M_{\mu \nu}\ .
\label{3daction}
\end{eqnarray}
\end{widetext} 
Here $e_i=\partial_0 a_i-\partial_i a_0$ and $b_i=\epsilon_{ijk}\partial_j a_k$ are the usual electric and magnetic fields associated with the gauge field $a_{\mu}$, while $f_{\mu} = (1/2) \epsilon_{\mu \nu \alpha \beta} \partial_{\nu} b_{\alpha \beta}$ is the dual field strength associated wiht the antisymmetric gauge field $b_{\mu \nu}$. In addition to the gauge symmetry under transformations $a_{\mu} \to a_{\mu} + \lambda$, this action is invariant under gauge symmetries of the second rank, $b_{\mu \nu} \to b_{\mu \nu} + \partial_{\mu} \chi_{\nu} -\partial_{\nu} \chi_{\mu}$, in which the gauge function itself is a vector. In 3D, vortices are one-dimensional extended objects and their world-surfaces are described by the two-index antisymmetric tensor $M_{\mu \nu}$. Cooper pairs, $Q_{\mu}$, and the related fluctuation number current $j_{\mu}= (\sqrt{2}/2\pi) f_{\mu}$ retain their point charge character. In 3D, $e_q$ is a dimensionless parameter, $e_{\mathrm q} = O(e)$, while $e_{\mathrm v}$ has the dimension of mass, $e_{\mathrm v}=O(1/\lambda)$, with $\lambda$ being the bulk London length of the material. The topological mass arising from the BF coupling\,\cite{bowick} maintains the same form as in 2D, $m_{\rs T}=\mu_{\rs P}  e_{\mathrm q}e_{\mathrm v}/\pi$. 

The derivation of the effective action for a superinsulator in 3D
follows exactly the same steps as in 2D, Supplementary note 2, Effective action for the superinsulator, with the result
\begin{widetext}
\begin{eqnarray}
S_{\rm eff}^{\rs SI} (A_{\mu}) = S_{\rm compact }^{\rs 3D}=
 {\gamma^2 \over 2\pi^2} \left\{ \sum_{x, i} 2v_{\mathrm c}\left[ 1- {\rm cos} \left( 2e\ell^2 \tilde F_{0i} \right)\right] +\sum_{x,i,j} {1\over v_{\mathrm c}} \left[ 1- {\rm cos} \left({ 2e\ell^2} \tilde F_{ij} \right) \right] \right\} \ .
\label{polya}
\end{eqnarray}
\end{widetext}
where $\tilde F_{\mu \nu} = k_{\mu \nu \alpha} A_{\alpha}$ is the 3D dual field strength ($k_{\mu \nu \alpha}$ being the 3D lattice BF term, see Methods, Lattice BF term). This is again a relativistic version of Polyakov's compact QED model, this time in 3D\,\cite{polyakov_original, polyakov}, with the relativistic ($v_c = 1$) string tension given by\,\cite{quevedo}
\begin{equation}
\sigma_{\rs{3D}} = {v_{\mathrm c}\over 64\pi \ell^2} K_0 \left( \frac{\sqrt{z}}{4\pi}\gamma  \right) \ ,
\label{tension3D}
\end{equation}
where $K_0$ is the McDonald function and $z$ is the monopole fugacity. Equations (\ref{comp}) and (\ref{polya}) are our key results, establishing an exact mapping between QCD and the physics of superinsulators, both in 2D and 3D. 

Finally, let us mention that, unlike in 2D, in 3D, the minimal coupling of charges to electromagnetism can be complemented by a topological coupling $\int d^4 x
\ i (\theta /8\pi \sqrt{2})\  \phi_{\mu \nu} F_{\mu \nu}$ of the vortex current $\phi_{\mu \nu} = (\sqrt{2}/2\pi) \epsilon_{\mu \nu \alpha \beta} \partial_{\alpha} a_{\beta}$ to the electromagentic field strength $F_{\mu \nu}$. This leads to an axion term\,\cite{axion} $S_{\rm axion} = \int d^4 x\  i(\theta/ 16\pi^2) F_{\mu \nu}\tilde F_{\mu \nu}$ in the electromagnetic effective action. This is a surface term, since the partition function ${\rm exp}\left( -S_{\rm axion} \right)$ is invariant under shifts $\theta \to \theta + 2\pi$. Time reversal, ${\cal T}$, maps $\theta \to -\theta$. So the only values of $\theta$ compatible with ${\cal T}$-invariance are $\theta = 0$ and $\theta = \pi$, modulo $2\pi$.
For $\theta = \pi$ the string becomes fermionic\,\cite{polyakov}, acquiring a topological contribution $(-1)^\nu$ in the partition function, where $\nu$ is the signed self-intersection number of the world-sheet in four-dimensional Euclidean space-time. 
The (relativistic) string tension changes to\,\cite{quevedo} 
\begin{equation}
\sigma_{\rs{3D}} = {v_{\mathrm c}\over  64\pi \ell^2} K_0 \left( {\sqrt{z} \over 16 \gamma} \right) 
\label{fermionic}
\end{equation}
Because the factor $\gamma$ is now in the denominator, the fermionic Cooper pair mesons are large also in the deep superinsulating region, where $\eta g \ll 1$ and $v =O(1)$.  

\begin{figure}
	\includegraphics[width=9cm]{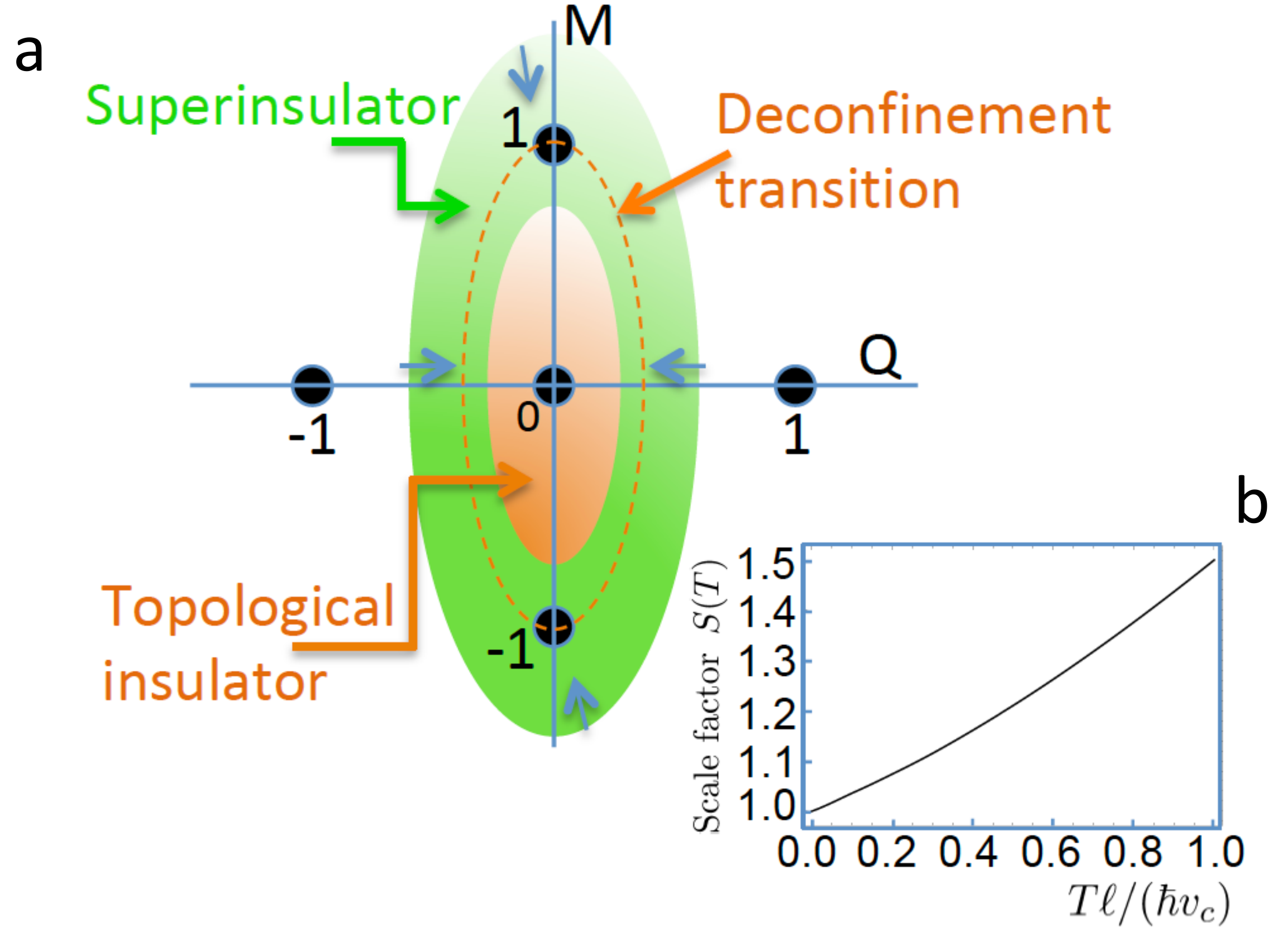}
	\caption{\label{fig:Fig4} \textbf{Deconfinement transition.} \textbf{a:} Finite temperature deconfinement transition from a superinsulator (magnetic numbers $M=\pm 1$ fall into the interior of the ellipse, while electric numbers $Q=\pm 1$ remain outside) to an insulator (no non-trivial quantum numbers fall within the ellipse). \textbf{b:} The finite-temperature scaling factor that determines the critical temperature for the superinsulator deconfinement transition, $v_{\mathrm c}$ is the light velocity in the material.}
\end{figure}

\subsection{Finite temperatures}
Now we turn to the finite temperature behavior and the deconfinement transition at which string confinement of Cooper pairs ceases to exist and the superinsulator transforms to a `conventional' insulator. This happens at the critical temperature $T_\mathrm{dc}$ where the linear tension of the string turns to zero. While it is known that, in 2D, $T_\mathrm{dc}\equiv T_{\rs BKT}$\,\cite{yaffe}, we can calculate $T_\mathrm{dc}$ straightforwardly as the temperature of disappearance of the vortex condensate. This is done in Methods, Finite temperature deconfinement transition, with the result that the superinsulator experiences a direct deconfinement transition to an insulating state at the critical deconfinement temperature determined by the equation $1/(g \eta) = S(T_\mathrm{dc})$ where the function $S(T)$ is derived by a geometric condition for the two competing condensations (see Supplementary note 3, Quantum phase transitions) and is shown in  Fig.\,\ref{fig:Fig4}. This equation uniquely determines the deconfinement temperature as a function of material parameters.

\subsection{Experimental implications}
To explore the far reaching experimental implications of the confining string theory of superinsulation note first 
that the deconfinement criticality depends on the space dimension\,\cite{VTF-DTV}. In 2D it coincides with that of the BKT transition\,\cite{yaffe}, and the resistance $R_{\rs{2D}}\propto\exp(b/\sqrt{|T/T_{\rs BKT}-1|})$. In 3D, instead, the resistance exhibits the so-called Vogel-Fulcher-Tamman (VFT) criticality, $R_{\rs{3D}}\propto\exp[b^{\prime}/|T/T_{\mathrm{dc}}-1|]$\,\cite{VTF-DTV}. Juxtaposing the critical behaviors of the NbTiN film, having a superconducting coherence length $\xi\gtrsim d$\,\cite{Mironov2018} and that of the InO film, where $\xi\ll d$\,\cite{Shahar}, one sees that the NbTiN film shows the BKT- while the InO film exhibits the VFT divergence, in compliance with our predictions about 3D superinsulation. 

The deconfinement transition can be realized as a quantum dynamical phase transition driven by an applied electric field $\bf{E}$ that would tear the electric strings. 
The threshold voltage, $V_{\mathrm t}\propto \sigma L$, corresponding to the pair-breaking critical current in superconductors, breaks down the neutral meson chains, and a strip of  `normal' insulator forms along the former string path,  carrying the current. This pretty much resembles the conventional dielectric breakdown where the electric field burns a conducting channel in otherwise insulating environment and triggers avalanche-like current jumps. The dielectric breakdown is usually accompanied by current noise. Such a noise has indeed been recently observed in InO films\,\cite{noise}. 
Experiments demonstrating the linear dependence of the threshold voltage on the sample size in films are still to come. Yet the evidence for linear confinement was provided by the analysis of the superinsulating behavior in the ultrathin TiN films\,\cite{Fistul2008}, which revealed that the magnetic field dependence of $V_{\mathrm t}$ is exactly that of the 1D Josephson ladder. 

In QCD, the flip side of the string confinement mechanism is asymptotic freedom, i.e. the unconstrained dynamics of quarks at spatial scales smaller than the string size\,\cite{gross}. While, strictly speaking, asymptotic freedom refers to the running of the dimensionless gauge coupling to zero in the ultraviolet limit, it can be viewed, from the string point of view, as the ``slackening" of the string so that quarks feel only weak short-range potentials at small scales. One would thus expect that, in superinsulators, asymptotic freedom, in this string sense, should map onto the unconstrained motion of the Cooper pairs at scales smaller than $d_{\rm string}$. The ratio of the string width to the string length is $w_{\rm string}/d_{\rm string} \propto (v_{\mathrm c}/\gamma^2) {\rm exp} (K \gamma^2/v_{\mathrm c})$ with $K$ being a numerical constant. For systems with small $K$ and large $\gamma^2$ this ratio is small. At scales $w_{\rm string} < r < d_{\rm string}$, Cooper pairs do not feel the string tension anymore but neither do they feel Coulomb interactions screened by the photon mass. 
Hence, one can expect a metallic-like low-temperature behavior of small samples that should have turned superinsulating had their size exceeded the typical dimension of the confining string, estimated as 
$d_{\rm string} \lesssim \hbar v_{\mathrm c} / k_{\rs B}T_{\rm BKT}$. Using the TiN films parameters\cite{vinokur2008superinsulator,vinokurAnnals} one obtains $d_{\rm string} \lesssim 60\,\mu$m. Remarkably, the study of the size dependence of superinsulating properties in TiN films\,\cite{Kalok2010} revealed that in films with lateral sizes, of 20\,$\mu$m and less, the insulating, thermally activated behavior saturates to the metallic one upon cooling to `superinsulating temperatures.' This complies with the expected asymptotic freedom behavior. However, it would be premature to take it as a conclusive evidence for the asymptotic freedom in superinsulators, and further experimental research is needed.  

\section*{Discussion}
We conclude by pointing out a close connection of the string confinement mechanism to concepts of many-body-localization (MBL)\,\cite{mbl}. It was recently shown that MBL-like behaviour may arise without exogenous disorder, 
due to strong interactions alone\,\cite{calabrese}, and that, in gauge theories, this is due to the endogenous disorder embodied by the mixing of superselection sectors\,\cite{scardicchio}, this process being 
identified as a transport-inhibiting mechanism due to confinement in the Schwinger model in 1D. In our setting, it is the Polyakov monopole instantons that play the role of endogenous spontaneous disorder. Accordingly, our summation over the instanton gas configurations acts as averaging over endogenous disorder\,\cite{polyakov_original, polyakov}. Importantly, the instanton formulation describes not only 1D, but the 2D and 3D physical dimensions as well. This spontaneous disordering mechanism has the same effect, that of mixing, in this case, the flux superselection sectors, leading to the survival of only the neutral charge sector as the physical state, while all other, charged states are localized on the string scale. Hence inhibition of the charge transport and the infinite resistance. The same confinement mechanism that prevents the observation of quarks is thus responsible for the absence of charged states and the infinite resistance in superinsulators.

\section*{Acknowledgements}
We are delighted to thank N.\,Nekrasov, M.\,Vasin, and Ya.\,Kopelevich for illuminating discussions. M.C.D. thanks CERN, where she completed this work, for kind hospitality. The work at Argonne (V.M.V.) was supported by the U.S. Department of Energy, Office of Science, Basic Energy Sciences, Materials Sciences and Engineering Division.

\section*{Appendix}
	\subsection{Lattice Chern-Simons operator}
	The formulation of a gauge-invariant lattice Chern-Simons term requires particular care. Following \cite{Diamantini1996} we introduce first the forward and backward derivatives and shift operators on a three-dimensional Euclidean lattice with sites denoted by $\{ x \}$, directions indicated by Greek letters and lattice spacing $\ell$, 
	\begin{eqnarray}
	d_{\mu} f(x) = {f(x +\ell \hat \mu) -f(x) \over \ell} \ , \qquad \qquad S_{\mu } f(x)= f(x +\ell \hat \mu) \ ,
	\nonumber \\
	\hat d_{\mu} f(x) = {f(x) -f(x +\ell \hat \mu)  \over \ell} \ , \qquad \qquad \hat S_{\mu } f(x)= f(x -\ell \hat \mu) \ .
	\label{shift}
	\end{eqnarray}
	Summation by parts on the lattice interchanges both the two derivatives (with a minus sign) and the two shift operators. Gauge transformations are defined by using the forward lattice derivative. In terms of these operators one can then define two lattice Chern-Simons terms
	\begin{equation}
	k_{\mu \nu} = S_{\mu} \epsilon_{\mu \alpha \nu} d_{\alpha} \ , \qquad \qquad \hat k_{\mu \nu} = \epsilon_{\mu \alpha \nu} \hat d_{\alpha} \hat S_{\nu} \ ,
	\label{latticecs}
	\end{equation}
	where no summation is implied over equal indices. Summation by parts on the lattice interchanges also these two operators (without any minus sign). Gauge invariance is then guaranteed by the relations
	\begin{equation}
	k_{\mu \alpha} d_{\nu} = \hat d_{\mu} k_{\alpha \nu} = 0 \ , \qquad \qquad \hat k_{\mu \nu} d_{\nu} = \hat d_{\mu} \hat k_{\mu \nu} = 0 \ .
	\label{gaugeinv}
	\end{equation}
	Note that the product of the two Chern-Simons terms gives the lattice Maxwell operator
	\begin{equation}
	k_{\mu \alpha} \hat k_{\alpha \nu} = \hat k_{\mu \alpha} k_{\alpha \nu} = -\delta_{\mu \nu} \nabla^2 +d_{\mu} \hat d_{\nu} \ ,
	\label{maxwell}
	\end{equation}
	where $\nabla^2 = \hat d_{\mu} d_{\mu}$ is  the 3D Laplace operator. 
	The discrete version of the mixed Chern-Simons gauge theory can thus be formulated as
	\begin{widetext}
	\begin{eqnarray}
	\nonumber S=\sum_x \  i  {\ell^3  \over \pi} a_{\mu} k_{\mu \nu} b_{\nu} 
	+{\ell^3 \over 2e^2_{\mathrm v} \mu_{\rs P}} f_0^2+ {\ell^3 \varepsilon_{\rs P} \over 2e^2_{\mathrm v}} f_i^2
	+{\ell^3 \over 2e^2_{\mathrm q}\mu_{\rs P}} g_0^2 + {\ell^3 \varepsilon_{\rs P} \over 2e^2_{\mathrm q}} g_i^2  
	+i\ell \sqrt{2} a_{\mu} Q_{\mu} +i\ell \sqrt{2} b_{\mu} M_{\mu}\ ,
	\label{nonrelac}
	\end{eqnarray}
	\end{widetext}
	where the discrete dual field strengths are given by
	\begin{equation}
	f_{\mu} = k_{\mu \nu} b_{\nu} \ , \qquad 
	g_{\mu} = k_{\mu \nu} a_{\nu} \ .
	\label{dual}
	\end{equation}
	As we show below, this action describes two massive modes with dispersion relation and mass given
	\begin{equation}
	E = \sqrt{m_{\rs T}^2 v_{\mathrm c}^4 + v_{\mathrm c}^2 k^2} \ , \qquad m_{\rs T}={\mu_{\rs P}  e_{\mathrm q} e_{\mathrm v} \over \pi} \ .
	\label{dispersion}
	\end{equation}
	where $v_{\mathrm c} = 1/\sqrt{\mu_{\rs P} \varepsilon_{\rs P}}$ is the light velocity in the medium. 
	This is the non-relativistic version of the celebrated Chern-Simons mass \cite{jackiw}. 
	\bigskip
	
	\subsection{Lattice BF operator}~\\
	The formulation of a discrete 3D lattice BF model \cite{bf} can be achieved along the same lines as in 2D. 
	Following \cite{Diamantini1996} we introduce the lattice $BF$ operators
	\begin{equation}
	k_{\mu \nu \rho} \equiv S_{\mu }\epsilon _{\mu \alpha
		\nu \rho } d _{\alpha } \qquad \qquad
	\hat k_{\mu \nu \rho} \equiv \epsilon _{\mu \nu \alpha \rho}
	\hat d_{\alpha }\hat S_{\rho } \ ,
	\label{kop}
	\end{equation}
	where 
	\begin{eqnarray}
	d_{\mu } f(x) &&\equiv {{f(x+\ell \hat \mu )-f(x)}\over
		\ell}\ ,\qquad S_{\mu }f(x) \equiv f(x+\ell \hat \mu )\ ,
	\nonumber \\
	\hat d_{\mu } f(x) &&\equiv {{f(x)-f(x-\ell \hat \mu )}\over \ell } \ ,
	\qquad \hat S_{\mu }f(x) \equiv f(x-\ell \hat \mu ) \ ,
	\label{shift}
	\end{eqnarray}
	are the forward and backward lattice derivative and shift operators, respectively. Summation by parts on the lattice interchanges both the two derivatives (with a minus sign) and the two shift operators; gauge transformations are defined using the forward lattice derivative. Also the 
	two lattice $BF$ operators are interchanged (no minus sign) upon summation. Moreover they are gauge invariant, in the sense that they obey the following equations:
	\begin{eqnarray}
	&&k_{\mu \nu \rho} d _{\nu } = k_{\mu \nu \rho}
	d _{\rho } = \hat d_{\mu } k_{\mu \nu \rho} = 0 \ ,
	\nonumber \\
	&&\hat k_{\mu \nu \rho }d _{\rho } = \hat d _{\mu }
	\hat k_{\mu \nu \rho} = \hat d_{\nu }
	\hat k_{\mu \nu \rho } = 0 \ .
	\label{gaugeinv}
	\end{eqnarray}
	Finally, they satisfy also the equations
	\begin{widetext}
	\begin{eqnarray}
	&&\hat k_{\mu \nu \rho} k_{\rho \lambda \omega} =
	-\left( \delta _{\mu \lambda} \delta_{\nu \omega} - \delta _{\mu \omega}
	\delta_{\nu \lambda } \right) \nabla^2  
	+ \left( \delta _{\mu \lambda }
	d_{\nu } \hat d_{\omega} - \delta _{\nu \lambda } d_{\mu }
	\hat d_{\omega } \right) 
	+ \left( \delta _{\nu \omega} d_{\mu }
	\hat d_{\lambda } - \delta _{\mu \omega} d_{\nu } \hat
	d_{\lambda } \right) \ ,
	\nonumber \\
	&&\hat k_{\mu \nu \rho} k_{\rho \nu \omega } = k_{\mu \nu \rho } \hat
	k_{\rho \nu \omega} = 2 \left( \delta _{\mu \omega } \nabla^2 - d_{\mu }
	\hat d_{\omega } \right) \ ,
	\label{maxwell}
	\end{eqnarray}
	where $\nabla^2 = \hat d_{\mu} d_{\mu}$ is the lattice Laplacian. 
	The Euclidean lattice BF model in 3D is then given by the action 
	\begin{eqnarray}
	\nonumber S = \sum_x i{\ell^4  \over \pi} a_{\mu} k_{\mu \alpha \beta} b_{\alpha \beta} +
	{\ell^4 \over 2e_{\mathrm q}^2 \mu_{\rs P}} b_i^2 + {\ell^4 \varepsilon_{\rs P} \over 2e_{\mathrm q}^2} e_i^2
	+{\ell^4 \over 2e_{\mathrm v}^2 }\mu_{\rs P} f_0^2 +{\ell^4 \epsilon_{\rs P}\over 2e_{\mathrm v}^2} f_i^2
	+i\ell \sqrt{2} a_{\mu } Q_{\mu }
	+i\ell ^2{\sqrt{2}\over 2} b_{\mu \nu} M_{\mu \nu} \ ,
	\label{ac}
	\end{eqnarray}
		\end{widetext}
	where the dual field strengths are now defined by 
	\begin{equation} 
	f_{\mu} = {1 \over 2} k_{\mu \nu \rho} b_{\nu \rho} \ , \qquad 
	\tilde f_{\mu \nu} = \hat k_{\mu \nu \rho} a_{\rho} \ ,
	\label{curr}
	\end{equation}
	and $e_i=d_0 a_i -d_i a_0$ and $b_i = \tilde f_{0i}$ are the usual electric and magnetic fields associated with the gauge field $a_{\mu}$. 
	The dispersion relation and mass remain identical to the 2D formulas. In this case they are the non-relativistic generalizations of the BF mass \cite{bowick}. 
	\bigskip

	\subsection{Finite Temperature Deconfinement Transition}~~\\
	In the field theory, the finite temperature $T$ is introduced by formulating the action on a Euclidean time of finite length $\beta =1/T$, with periodic boundary conditions (we have reabsorbed the Boltzmann constant into the temperature). If the original field theory model is defined on a Euclidean lattice of spacing $\ell$, then $\beta $ is quantized in integer multiples of $\ell/v_{\mathrm c}$. This representation of the finite-temperature field theory holds as long as $v_{\mathrm c}\beta \gg \ell$, or, equivalently, if the temperature is much lower than the UV cutoff, $T \ll v_{\mathrm c}/\ell$, as expected. 
	Because of the lattice structure, energies are defined only within a Brillouin zone of length $2v_{\mathrm c}\pi/\ell$,
	due to the periodic boundary condition in the Euclidean time direction, however the energy $k^0$ must be also quantized in the integer multiples of $2\pi/\beta$. This gives 
	\begin{equation}
	\int_{0}^{2\pi v_{\mathrm c}\over \ell} d k^0 f \left( k^0 \right)  \to \sum_{n=0}^{n=b} {2\pi \over \beta} f\left( {2\pi v_{\mathrm c} n\over b\ell } \right) \ ,
	\label{finite}
	\end{equation}
	where $\beta = b \ell/v_{\mathrm c}$ and the factor within the sum represents the density of states. The integers $n$ in the summation are known as Matsubara frequencies. Typically, however momenta integral are defined over the fundamental Brillouin zone $[ -\pi v_{\mathrm c}/\ell, \pi v_{\mathrm c}/\ell ]$, rather then $[0,2\pi v_{\mathrm c} /\ell]$. The corresponding finite temperature expression can be readily obtained from (\ref{finite}) by the shift $k^0 \to k^0 -\pi v_{\mathrm c}/\ell$, 
	\begin{equation}
	\int_{-\pi v_{\mathrm c} \over \ell }^{\pi v_{\mathrm c}\over \ell} d k^0 f \left( k^0 \right)  \to \sum_{k=-b}^{k=b} {\pi  \over \beta} f\left( {\pi v_{\mathrm c} k\over b\ell } \right) \ ,
	\label{finitetemp}
	\end{equation}
	where $k = 2n-b$ and thus correspondingly, the density of states must be divided by a factor 2. 
	
	The finite temperature $T>0$ affects primarily the parameter $\eta$ (see Supplementary note 3, Quantum phase structure) via the coefficient $G(m\ell v_{\mathrm c})$.
	At the zero temperature this is given by
	\begin{equation}
	G(m\ell v_{\mathrm c}) = {1\over (2\pi)^4} \int_{-\pi}^{\pi} d^4k { 1 \over (m\ell v_{\mathrm c})^2 + \sum_{i=0}^3 4\  {\rm sin} \left( {k^i\over 2} \right)^2 } \ .
	\label{gzero}
	\end{equation}
	At finite temperatures it has to be modified according to (\ref{finitetemp}), 
	\begin{widetext}
	\begin{equation}
	G(m\ell v_{\mathrm c}, T) ={1\over (2\pi)^4}
	\sum_{k=-b}^{k=+b} {\pi \over b} \int_{-\pi}^{\pi} { dk^1 dk^2 dk^3 \over (m\ell v_{\mathrm c})^2 + 4\ {\rm sin} \left( {\pi k \over 2b} \right)^2 
		+ \sum_{i=1}^3 4\  {\rm sin} \left( {k^i\over 2} \right)^2 } \ ,
	\label{gfinite}
	\end{equation}
	\end{widetext}
	where $T=v_{\mathrm c}/b\ell$. 
	As we have verified over 3 orders of magnitude ($m\ell v_{\mathrm c}=0.001$ to $m\ell v_{\mathrm c}=1$) the ratio $S(T) = G(m\ell v_{\mathrm c}, T)/G(m\ell v_{\mathrm c})$ does not depend on the parameter $m\ell v_{\mathrm c}$ but is rather a function of the temperature alone. As a consequence, $\eta$ and the semiaxes of the ellipse determining the phase structure, see Supplementary note 3, Supplementary Equations (33), scale with the inverse of the function $S(T)$. This means that with the increasing temperature the whole ellipse shrinks by the scale factor $S(T)$. Magnetic quantum numbers $M=\pm 1$ that are within the ellipse at $T=0$, will exit its interior at some critical temperature defined by the condition
	\begin{equation}
	{1\over g\eta} = S(T_{\mathrm c}) \,,
	\label{crit}
	\end{equation}
	assuming that the quantity on the left-hand side is larger than one (i.e. there is a superinsulator at $T=0$). Since the magnetic semiaxis is always longer and thus no electric quantum numbers may appear within the ellipse interior when the magnetic ones have fallen outside, 
	the superinsulator experiences a direct deconfinement transition into a topological insulator at $T=T_{\mathrm c}$. Correspondingly, superconductors undergo a phase transition to topological insulators at $\tilde T_{\mathrm c}$ defined by 
	\begin{equation}
	{g\over \eta} = S(\tilde{T}_{\mathrm c})\,.
	\label{crittilde}
	\end{equation}
	
\section*{Supplementary Notes}

\subsection*{Gauge theory of Josephson junction arrays (JJA)} 
In this note we demonstrate that the general topological action of the Cooper pairs-vortex system in two dimensions, given by Eq.\,(3) of the main text, naturally arises for the lateral Josephson junction array (JJA). Our starting point is the coupled Coulomb gas description of\,\cite{fazio} (Eq.\,(31) there). We will consider first the continuum formulation of the coupled Coulomb gas, which  is given by the Euclidean action
\begin{eqnarray}
S &=& \int d^3 x \left[ 4E_{\rs C}C\ \rho_{\rs Q} {1\over -\nabla^2} \rho_{\rs Q} + 2\pi^2 E_{\rs J} \ \rho_{\rs V} {1\over -\nabla^2} \rho_{\rs V} +{1\over 2E_{\rs J}} \dot \rho_{\rs Q} {1\over -\nabla^2} \dot \rho_{\rs Q} \right]
\nonumber \\
&+ &i \int dt \int d^2{\bf x} \int d^2 {\bf y} \left[ \rho_{\rs Q} (t,{\bf x}) \Theta ({\bf x}-{\bf y}) \dot \rho_{\rs V} (t,{\bf y})\right] \ ,
\label{initial}
\end{eqnarray}
where $E_{\rs J}$ is the Josephson coupling, $E_{\rs C} = 2e^2/C$ (with $C$ the junction capacitance) is the charging energy of the array and we have used
\begin{equation}
- {\rm ln} |{\bf x}| ={2\pi \over -\nabla^2} \delta ({\bf x}) \ ,
\label{log}
\end{equation}
and
\begin{equation}
\Theta ({\bf x}) = {\rm arctan} \left( {x_2\over x_1} \right) \ .
\label{arctan}
\end{equation}
As usual in statistical field theory, this action plays the same role as the Hamiltonian of a 3D statistical mechanics model, with the relevant coupling constant taking the role of temperature. 
The first two terms in the action represent the two Coulomb gases for charges with density $\rho_{\rs Q}$ and vortices with density $\rho_{\rs V}$, the third is a kinetic term for the charges and the final term represents the Aharonov-Bohm topological interaction between charges and vortices. The only term which breaks perfect duality between charges and vortices in this expression is the kinetic term for charges, which encodes the Josephson currents. The self-dual approximation, originally introduced in\,\cite{Diamantini1996}, consists in adding a corresponding kinetic term for vortices and modifying thus the action as follows 
\begin{eqnarray}
S &&=\int d^3 x \ 4E_{\rs C} \ \rho_{\rs Q} {1\over -\nabla^2} \rho_{\rs Q} + 2\pi^2 E_{\rs J} \ \rho_{\rs V} {1\over -\nabla^2} \rho_{\rs } 
\nonumber \\
&&+ \int d^3 x \ {1\over 2E_{\rs J}} \dot \rho_{\rs Q} {1\over -\nabla^2} \dot \rho_{\rs Q} + {\pi^2 \over 4E_{\rs C}} \dot \rho_{\rs V} {1\over -\nabla^2} \dot \rho_{\rs V} 
\nonumber \\
&&+ i \int dt \int d^2{\bf x} \int d^2 {\bf y} \left[ \rho_{\rs Q} (t,{\bf x}) \Theta ({\bf x}-{\bf y}) \dot \rho_{\rs V} (t,{\bf y})\right] \ ,
\label{initial}
\end{eqnarray}
Note that this is a harmless modification, since such a kinetic term for vortices is anyhow radiatively induced by integration over the charge dynamics, as is derived, e.g. in Eq.\,(34) of\,\cite{fazio}.  

In order to proceed with the gauge theory derivation we consider the action formulated in the Minkowski space, with the only change of missing ``i" in the interaction term between the charges and vortices,
\begin{eqnarray}
S_{\rs M} &&= \int d^3 x \ 4E_{\rs C} \ \rho_{\rs Q} {1\over -\nabla^2} \rho_{\rs Q} + 2\pi^2 E_{\rs J} \ \rho_{\rs V} {1\over -\nabla^2} \rho_{\rs V} 
\nonumber \\
&&+\int d^3 x \ {1\over 2E_{\rs J}} \dot \rho_{\rs Q} {1\over -\nabla^2} \dot \rho_{\rs Q} + {\pi^2 \over 4E_{\rs C}} \dot \rho_{\rs V} {1\over -\nabla^2} \dot \rho_{\rs V} 
\nonumber \\
&&+\int dt \int d^2{\bf x} \int d^2 {\bf y} \left[ \rho_{\rs Q} (t,{\bf x}) \Theta ({\bf x}-{\bf y}) \dot \rho_{\rs V} (t,{\bf y})\right] \ ,
\label{mink}
\end{eqnarray}
Now we combine the two independent variables encoded in the charge density and its time derivative in a single dual gauge field strength by introducing for charges and vortices two fictitious gauge fields $a_{\mu}$ and $b_{\mu}$, 
\begin{eqnarray}
f^{i} = \epsilon^{ij} \partial_j b_0 - \epsilon^{ij} \dot b_{j} \ ,
\nonumber \\
g^{i} = \epsilon^{ij} \partial_j a_0 -\epsilon^{ij} \dot a_j \ .
\label{unify}
\end{eqnarray}
In this representation we take the spatial gauge fields $a_i$ and $b_j$ as transverse, $a_i = \epsilon^{ij}\partial_j \eta$ and $b_i = \epsilon^{ij}\partial_j \chi$, since a longitudinal part can be reabsorbed by a redefinition of $a_0$ and $b_0$. Rewriting the action as
\begin{eqnarray} 
S_{\rs M} &=& \int d^3 x \left[ {1\over 2e^2_{\mathrm v}}  f^i f^i + {1\over 2e^2_{\mathrm q}} g^i g^i + \sqrt{2} \ a_0 \rho_{\rs Q} + \sqrt{2} \ b_0 \rho_{\rs V} \right]
\nonumber \\
&+ &\int dt \int d^2{\bf x} \int d^2 {\bf y} \left[ \rho_{\rs Q} (t,{\bf x}) \Theta ({\bf x}-{\bf y}) \dot \rho_{\rs V} (t,{\bf y})\right] \ ,
\label{modaction}
\end{eqnarray}
with
\begin{eqnarray}
e^2_{\mathrm v} &=& 2\pi^2 E_{\rs J} \ ,
\nonumber \\
e^2_{\mathrm q} &= & 4E_{\rs C} \ ,
\label{cconst}
\end{eqnarray}
and realizing that
\begin{eqnarray}
\int d^3 x \ {1\over 2e^2_{\mathrm v}}  f^i f^i &=& \int d^3x \left[  {1\over 2e^2_{\mathrm v}} b_0 (-\nabla^2) b_0 + {1\over 2e^2_{\mathrm v}} \dot \chi ( -\nabla^2) \dot \chi\right] \ ,
\nonumber \\
\int d^3 x \ {1\over 2e^2_q}  g^i g^i &= &\int d^3x \left[  {1\over 2e^2_{\mathrm q}} a_0 (-\nabla^2) a_0 + {1\over 2e^2_{\mathrm q}} \dot \eta ( -\nabla^2) \dot \eta\right] \ ,
\label{coulomb}
\end{eqnarray}
one obtains the two Coulomb interactions for charges and vortices by eliminating the non-dynamical Lagrange multipliers $a_0$ and $b_0$ after having solved the Gauss law constraints
\begin{eqnarray} 
\nabla^2 b_0 &= &\sqrt{2} e^2_v \ \rho_{\rs V} \ ,
\nonumber \\
\nabla^2 a_0 &=& \sqrt{2} e^2_q \ \rho_{\rs Q} \ .
\label{gauss}
\end{eqnarray}

We can now pack also the charges into a complete gauge theory formulation by rewriting 
\begin{eqnarray}
\rho_{\rs Q} &=& (\sqrt{2}/ 2\pi ) f^0 \ ,
\nonumber \\
\rho_{\rs V} &=& (\sqrt{2}/ 2\pi ) g^0 \ ,
\label{cv}
\end{eqnarray}
where $f^0$ and $g^0$ are the time components of three-dimensional dual field strengths 
\begin{eqnarray}
f^{\mu} &&= {1\over 2} \epsilon^{\mu \nu \alpha} f_{\nu \alpha}  = \epsilon^{\mu \nu \alpha} \partial_{\nu} b_{\alpha} \ ,
\nonumber \\
g^{\mu} &&= {1\over 2} \epsilon^{\mu \nu \alpha} g_{\nu \alpha}  = \epsilon^{\mu \nu \alpha} \partial_{\nu} a_{\alpha} \ .
\label{dualgauge}
\end{eqnarray}
In the gauge theory formulation these represent the conserved charge and vortex currents
\begin{eqnarray}
j^{\mu} &&= {\sqrt{2}\over 2\pi} f^{\mu} \ ,
\nonumber \\
\phi^{\mu} &&= {\sqrt{2}\over 2\pi} g^{\mu} \ .
\label{currents}
\end{eqnarray}
The Bianchi identities for $\partial_{\mu }f^{\mu }$ and $\partial_{\mu }g^{\mu }$ then yield
\begin{eqnarray}
\dot \rho_{\rs Q} &=& { \sqrt{2}\over 2\pi} (-\nabla^2) \dot \chi \ ,
\nonumber \\
\dot \rho_{\rs V} &=& { \sqrt{2}\over 2\pi} (-\nabla^2) \dot \eta \ .
\label{dots}
\end{eqnarray}
Substituting these expressions in\,(\ref{modaction}) via\,(\ref{coulomb}) one obtains also the kinetic terms in\,(\ref{mink}). 

Finally, using 
\begin{eqnarray}
\rho_{\rs Q} &=& {\sqrt{2}\over 2\pi } f^0 = {\sqrt{2}\over 2\pi } \epsilon^{ij}\partial_i b_j \ ,
\nonumber \\
\partial_i \eta &=& -\epsilon^{ij} a_j \ ,
\nonumber \\
\partial_i \chi &= -&\epsilon^{ij} b_j \ ,
\label{help}
\end{eqnarray}
and the identity 
\begin{equation}
\epsilon^{ij} {\partial_j \over \nabla^2} \delta^2({\bf x}) = -{1\over 2\pi} \partial_i \ {\rm arctan} \left( x^2 \over x^1\right) \ ,
\label{relation}
\end{equation}
one can transform the last topological term in\,(\ref{mink}) into $\int d^3 x \ (-1/\pi) b_i \epsilon^{ij} \dot a_j$ which can be recombined with the two Lagrange multipliers into the Chern-Simons term \cite{jackiw} $(1/\pi ) a_{\mu} \epsilon^{\mu \nu \alpha} \partial_{\nu} b_{\alpha}$. This gives the total action
\begin{equation} 
S = \int d^3 x \left[ {1\over 2e^2_{\mathrm v}}  f^i f^i + {1\over 2e^2_{\mathrm q}} g^i g^i +i  {1\over \pi} a_{\mu} \epsilon^{\mu \nu \alpha} \partial_{\nu} b_{\alpha} \right]\ ,
\label{action}
\end{equation}
where we have rotated back to the original Euclidean space.

\subsection*{Effective action for the superinsulator}
To explore the implications of vortex condensation we shall focus on the 3D case and consider the SIT effective action in its lattice formulation. The corresponding superinsulator partition fuction is given by 
\begin{equation}
Z_{\rs SI} = \int_{a_{\mu}, b_{\mu \nu}} {\cal D} a_{\mu} {\cal D} b_{\mu} \sum_{\{ M_{\mu \nu} \}}\ {\rm e}^{-S\left( a_{\mu}, b_{\mu  \nu}\right)}\  {\rm e}^{i\ell ^2{\sqrt{2}\over 2} b_{\mu \nu}M_{\mu \nu}} \ ,
\label{parfct}
\end{equation}
where $S\left( a_{\mu}, b_{\mu  \nu}\right)$ represents the pure gauge part of the action and we have set $Q_{\mu} =0$ since charges are dilute in this phase. The integers $M_{\mu \nu}$ are the lattice representation of the world-surfaces spanned by the one-dimensional vortices. For closed vortices, we can express these integers as 
in the form $M_{\mu \nu} = \ell \hat k_{\mu \nu \alpha} n_{\alpha}$ with $n_{\alpha} \in \mathbb{Z}$,
\begin{equation}
Z_{\rs SI} = \int_{a_{\mu}, b_{\mu}} {\cal D} a_{\mu} {\cal D} b_{\mu} \sum_{\{ n_{\mu} \}}\ {\rm e}^{-S\left( a_{\mu}, b_{\mu  \nu}\right)}\  {\rm e}^{i\ell ^3{\sqrt{2}\over 2} n_{\mu} k_{\mu \alpha \beta} b_{\alpha \beta}} \ .
\label{parfct2}
\end{equation}
Finally, we can turn the sum over $n_{\mu}$ into an integral by the usual Poisson formula
\begin{equation}
\sum_{n_{\mu}} f\left( n_{\mu} \right) = \sum_{k_{\mu}} \int dn_{\mu} f\left( n_{\mu} \right) {\rm e}^{i2\pi n_{\mu}k_{\mu}} \ ,
\label{poisson}
\end{equation}
where the new integer link variables  $\{ k_{\mu} \}$ must satisfy $\hat d_{\mu}k_{\mu}=0$ in order to guarantee gauge invariance under transformations $n_{\mu} \to n_{\mu} + \ell d_{\mu} t$. In principle, this gauge invariance should be gauge fixed in (\ref{parfct2}) by introducing, e.g. a quadratic term in $n_{\mu}$. Removing the gauge fixing in the final result gives back the same result obtained by naive integration, 
\begin{equation}
Z_{\rs SI} = \int_{a_{\mu}, b_{\mu}} {\cal D} a_{\mu} {\cal D} b_{\mu} \sum_{\{ k_{\mu} \}}\ {\rm e}^{-S\left( a_{\mu}, b_{\mu  \nu}\right)}\  \delta \left( \ell^3 k_{\mu \alpha \beta} b_{\alpha \beta} - \sqrt{2} 2\pi k_{\mu}\right)  \ .
\label{parfct3}
\end{equation}
The important consequence of this result is that gauge fields $\ell^2 b_{\mu \nu}$ become quantized in units of $(2\pi \sqrt{2})$ so that the original gauge symmetry with gauge group $\mathbb{R}$ is broken down to $\mathbb{Z}$, with allowed gauge functions $\lambda_{\mu} = (2\pi \sqrt{2}) i_{\mu}$, with $i_{\mu} \in \mathbb{Z}$. This means that magnetic monopoles $\hat d_{\mu} M_{\mu \nu}$ are allowed at the boundaries of magnetic surfaces, since ${\rm exp}(-S)$ is invariant under the restricted gauge transformations. The magnetic world-surfaces are actually open, reflecting the fact that the vortex number is not conserved in a condensate. In 3D, the magnetic monopoles are particles, in 2D, they are tunneling events at the end of vortex world-lines, i.e. instantons\,\cite{polyakov}. 
We can now unravel the nature of the superinsulator by adding to\,(\ref{parfct}) the minimal coupling $i\ell^4 2e A_{\mu} j_{\mu}$ of Cooper pairs to the external electromagnetic field $A_{\mu}$, where $j_{\mu } = {\sqrt{2}\over 2\pi} h_{\mu} = {\sqrt{2}\over 2\pi}\hat k_{\mu \nu \rho} b_{\nu \rho}$ and by computing the effective action $S_{\rm eff}^{\rm SI} (A_{\mu})$. 
Integrating out the gauge fields and retaining only the dominant self-interactions we obtain 
\begin{eqnarray}
&&{\rm exp}\left( -S_{\rm eff}^{\rm SI}\left(A_{\mu}\right) \right) =\nonumber\\
&&=\,\sum_{M_{\mu \nu}} {\rm exp} \left\{ -\gamma^2 \left[ \sum_{x, i} 2v_{\mathrm c}   \left( {2e \ell^2 \tilde F_{0 i}\over 2\pi} -M_{0i} \right)^2 + \sum_{x, i,j} {1\over v_{\mathrm c}} \left( {2e \ell^2 \tilde F_{ij}\over  2\pi} -M_{ij} \right)^2 \right] \right\} \,,
\label{villain}
\end{eqnarray}
where  
\begin{equation}
\gamma^2 = {g\eta \mu_{\rs A} \over v_{\mathrm c}} \ ,
\label{gamma}
\end{equation}
and $\tilde F_{\mu \nu} = \hat k_{\mu \nu \alpha} A_{\alpha}$ is the dual electromagentic field strength, $g$ and $\eta $ are the two parameters governing the phase structure and $\mu_{\rs A} =O(1)$ is a numerical parameter related to the entropy of surfaces. 
Since, as shown above, the condensing magnetic excitations $M_{\mu \nu}$ over which we have to sum in the partition function are unconstrained integers, allowing for magnetic monopoles, this 
is nothing else than a Villain formulation of the non-relativistic version of the famed Polyakov compact QED model\,\cite{polyakov}:
\begin{eqnarray}
S_{\rm eff}^{\rm SI} \left( A_{\mu}\right) &&=
{\gamma^2 \over 2\pi^2} \sum_{x, i} 2v_{\mathrm c}\left[ 1- {\rm cos} \left( 2e\ell^2 \tilde F_{0i} \right)\right] 
\nonumber \\
&&+ {\gamma^2 \over 2\pi^2}\sum_{x,i,j} {1\over v_{\mathrm c}} \left[ 1- {\rm cos} \left( 2e{\ell^2 } \tilde F_{ij} \right) \right] \ .
\label{polyakov}
\end{eqnarray}
In 2D, calculations follow exactly the same lines and give the analogous result with the entropy of surfaces $\mu_{\rs A}$ substituted by the entropy of lines $\mu \approx {\mathrm{ln}}\,5$. 
\bigskip

\subsection*{Quantum phase structure}
In this Supplementary Note we derive the structure of the phases in the vicinity of the SIT so that one could posit the superinsulating state with respect to superconducting and Bose metal phases in the phase diagram and introduce the graphic ellipsoid technique for determining the phase that realizes at the given values of the parameters. This technique is used to calculate the temperature of the deconfinement transition in the main text.  
We start with integrating out the gauge fields in Eq.\,(12) of Appendix of the main text in order to obtain an effective Euclidean action for point charges and vortices alone. The real part of this action, the one that enters the determination of the phase structure, is given by
\begin{eqnarray}
&&S_{\rm top}^{\rm real} = v_{\mathrm c}^2 {\sqrt{\mu_{\rs P}} \over \sqrt{\varepsilon_{\rs P}}\ell} \times
\nonumber \\
&&\sum_x  \left[  Q_0 {e^2_{\mathrm q}v_{\mathrm c}\over v_{\mathrm c}^4  m_{\rs T}^2- d_0\hat d_0 -v_{\mathrm c}^2\nabla_2^2} Q_0 + Q_i {(e^2_{\mathrm q}/ v_{\mathrm c})\ \delta_{ij} \over v_{\mathrm c}^4  m_{\rs T}^2-d_0\hat d_0 -v_{\mathrm c}^2 \nabla_2^2} Q_j\right]  +
\nonumber \\
&&\sum_x  \left[  M_0 {e^2_{\mathrm v}v_{\mathrm c}\over v_{\mathrm c}^4  m_{\rs T}^2-d_0\hat d_0 -v_{\mathrm c}^2 \nabla_2^2} M_0 + M_i {(e^2_{\mathrm v}/v_{\mathrm c})\ \delta_{ij} \over v_{\mathrm c}^4  m_{\rs T}^2-d_0\hat d_0 -v_{\mathrm c}^2\nabla_2^2} M_j\right]  \ ,
\label{topnonrela}
\end{eqnarray}
where $\nabla_2^2$ denotes the 2D Laplacian and $Q_{\mu}$ and $M_{\mu}$ are the integer link variables describing the charge and vortex world-lines, respectively. 

In order to proceed we follow the standard arguments of\,\cite{kogut} to retain only the self-interaction terms in\,(\ref{topnonrela}). Near the transition where large loops and long strings condense, the typical configurations of the fields $Q_{\mu}$ and $M_{\mu}$ are very rare.  Therefore, one may expect that the forces on every bond due to its neighbours (both on the same loop and on other ones) cancel out. Consider a closed string made of $N$ bonds, with integer quantum numbers $Q_{\mu } = Q$ and $M_{\mu} = M$ on all the lattice bonds forming the string and zero elsewhere. This corresponds to a fluctuation in which a charge-anticharge or vortex-antivortex pair is created from the vacuum, lives for a ``time" proportional to its length in the 0 direction and is then annihilated in the vacuum again. We are interested in long-living fluctuations, in which the dominant contribution to the action comes from the ``time" terms, first and third terms in (\ref{topnonrela}). These fluctuations can be assigned an energy (equivalent to Euclidean action in statistical field theory) 
\begin{equation}
S_{\rm top} = \pi m_{\rs T}\ell v_{\mathrm c}^2 G(m_{\rs T}\ell v_{\mathrm c}) \left[ {e_{\mathrm q}\over e_{\mathrm v}} \ Q^2 + {e_{\mathrm v}\over e_{\mathrm q}} \ M^2 \right] N \ ,
\label {ener}
\end{equation}
where $G(m_{\rs T}\ell v_{\mathrm c})$ is the diagonal element of the lattice kernel $G(m_{\rs T}\ell v_{\mathrm c}, x-y)$ representing the inverse of the operator $(\ell^2/v_{\mathrm c}^2) (m_{\rs T}^2v_{\mathrm c}^4-d_0\hat d_0 -v_{\mathrm c}^2\nabla_2^2)$. The kernel $G(m_{\rs T}\ell v_{\mathrm c}, x)$ is defined by the equation 
\begin{equation}
(\ell^2/v_{\mathrm c}^2) (m_{\rs T}^2v_{\mathrm c}^4-d_0\hat d_0 -v_{\mathrm c}^2\nabla_2^2) \ G(m_{\rs T}\ell v_{\mathrm c}, x) = \delta_{x,0} \ .
\label{green}
\end{equation}
Defining the Fourier transform $G(m_{\rs T}\ell v_{\mathrm c}, x) = \int _{-\pi v_{\mathrm c}/\ell}^{+\pi v_{\mathrm c}/\ell} dk_0 \int_{-\pi/\ell }^{+\pi/\ell} d^2k \ G(m_{\rs T}\ell v_{\mathrm c}, k) \ {\rm exp} (ik \cdot x)$ we obtain 
\begin{eqnarray}
\int _{-\pi v_{\mathrm c}/\ell}^{+\pi v_{\mathrm c}/\ell}&&dk_0 \int_{-\pi/\ell }^{+\pi/\ell} d^2k\ G(m_T\ell v_c, k) \ (\ell^2/v_c^2) (m_T^2v_c^4-d_0\hat d_0 -v_c^2\nabla_2^2)  {\rm e}^{ik \cdot x}
\nonumber \\
&&= {1\over (2\pi)^3} \int_{\pi}^{\pi} d^3k \ {\rm e}^{ik \cdot x} \ .
\label{appl}
\end{eqnarray}
Applying finally the finite difference operator $(\ell^2/v_c^2) (m_T^2v_c^4-d_0\hat d_0 -v_c^2\nabla_2^2)$ to the exponential in the Fourier transform and rescaling momenta gives the final result 
\begin{equation}
G(m_{\rs T}\ell v_{\mathrm c}) = {1\over (2\pi)^3} \int_{-\pi}^{\pi} d^3k { 1 \over (m_{\rs T}\ell v_{\mathrm c})^2 +  \sum_{i=0}^2 4\  {\rm sin} \left( {k^i\over 2} \right)^2 } \ .
\label{greenfinal}
\end{equation}

The string entropy, however is also proportional to their length, being given by $\mu N$ with $\mu \approx {\ln}\,5$ since at each step the non-backtracking strings can choose among 5 possible directions on how to continue. One can thus assign the free energy 
\begin{equation}
F = \pi m_{\rs T}\ell v_{\mathrm c}^2 G(m_{\rs T}\ell v_{\mathrm c}) \left[ {e_{\mathrm q}\over e_{\mathrm v}} \ Q^2 + {e_{\mathrm v}\over e_{\mathrm q}} \ M^2 - {1\over \eta} \right] N  \ ,
\label{freeenergy}
\end{equation}
to a string of length $L=\ell N$ carrying electric and magnetic quantum numbers $Q$ and $M$, respectively. Here we have introduced the dimensionless parameter
\begin{equation}
\eta = {\pi m_{\rs T}\ell v_{\mathrm c}^2 G(m_{\rs T}\ell v_{\mathrm c}) \over \mu}\ ,
\label{eta}
\end{equation}
which, together with the ratio $g=e_{\mathrm v}/e_{\mathrm q}$ fully determines the quantum phase structure, as we now show. 

The ground state of the quantum model is found by minimizing its free energy as a function of $N$. When the energy term in\,(\ref{freeenergy}) dominates, the free energy is positive and consequently minimized by short closed loop configurations. When, instead, the entropy dominates, the free energy is negative and minimized by large strings and long closed loops. The condition for condensation of long strings with integer quantum numbers  $Q$ and $M$ is thus given by
\begin{equation}
\eta {e_{\mathrm q}\over e_{\mathrm v}} Q^2 + \eta {e_{\mathrm v}\over e_{\mathrm q}} M^2 < 1 \ .
\label{ellipse}
\end{equation}
If two or more condensations are allowed, one has to choose the one with the lowest free energy. 
This condition describes the interior of an ellipse with the semiaxes 
\begin{eqnarray}
r_{\rs Q} = \sqrt{{e_{\mathrm v}\over e_{\mathrm q}} {1\over \eta}} \ ,
\nonumber \\
r_{\rs M}= \sqrt{{e_{\mathrm q}\over e_{\mathrm v}} {1\over \eta}} \ ,
\label{semiaxes}
\end{eqnarray}
on a square lattice of integer electric and magnetic charges. The phase diagram is consequently found by simply recording which integer charges lie within the ellipse when the semi-axes are varied,
\begin{eqnarray}
\eta &&< 1 \to 
\begin{cases}
{g}>1\ , {\rm electric \ condensation \ = \ superconductor} \ ,\\
{g}<1\ , {\rm magnetic \ condensation \ = \ superinsulator} \ ,\\
\end{cases}
\nonumber \\
\eta &&> 1 \to
\begin{cases}
{g} > {\eta}
\ , {\rm electric \ condensation \ = \ superconductor} \ ,\\
{\eta}
>{g}>{1\over \eta} \ , {\rm no \ condensation \ = \ quantum \ metal} \ ,\\
{g} < {1\over \eta} \ , {\rm magnetic \ condensation \ = \ superinsulator }\ ,\\
\end{cases}
\nonumber
\end{eqnarray}
This approach immediately shows that the possible intermediate phase opening up for $\eta > 1$, typically called a Bose (or quantum) metal\,\cite{qm}, is nothing else than a bosonic topological insulator\,\cite{hasan}, with a topological long-distance effective action\,\cite{moore} (mixed Chern-Simons in 2D, BF in 3D). 
Note that, for $\eta < 1$, there is a coexistence region for electric and magnetic charges when 
$g$ lies between $\eta $ and $1/\eta$. This indicates that the direct transition from a superinsulator to a superconductor is actually a first-order transition with metastable coexistence of phases in a region around it. 
The parameter $\eta$, determining the possible appearance of an intermediate topological insulator phase, acquires a particularly telling form in the vicinity of the SIT where $e_{\mathrm q} \approx e_{\mathrm v}$. In this case 
\begin{equation}
\eta = {1\over \alpha} {(v_{\mathrm c}/c)^2\over k} {\pi^2\over \mu} G\left( {(v_{\mathrm c}/c)\over \alpha k} \right) \ ,
\label{eta}
\end{equation}
where we have reinstated $\hbar$ and $c$, $\alpha = e^2/(\hbar c) \approx 1/137$ is the fine structure constant and $k=\lambda_{\perp}/\xi$ is the Landau parameter of the superconducting film.

\end{document}